\documentstyle[aps,twocolumn,epsf]{revtex}

\def\be{\begin{equation}}
\def\ee{\end{equation}}
\def\bea{\begin{eqnarray}}
\def\eea{\end{eqnarray}}
\def\sumint{\,\int\hspace{-1.3em}\sum}
\def\gsim{\raisebox{-0.5ex}{\mbox{ $\stackrel>\sim$ }}}
\def\pa{\partial}

\begin{document}

\title{Quasiparticle description of strongly coupled plasmas}
\author{Andr\'e Peshier}
\address{
 Forschungszentrum Rossendorf, PF 510119, 01314 Dresden, Germany,
 \\
 Institut f\"ur Theoretische Physik, Technische Universit\"at
 Dresden, 01062 Dresden, Germany}
\date{November 5, 1998}

\maketitle

\begin{abstract}
 The extrapolation of truncated perturbative series to the large coupling
 regime is not reliable since the accuracy of the calculations cannot be
 improved systematically by a finite set of higher order terms.
 Instead, a nonperturbative approach is proposed here to describe the
 thermodynamics of strongly coupled plasmas.
 Within the $\phi^4$ theory, it is derived from the Luttinger-Ward theorem
 and amounts to a quasiparticle description.
 The approach is extended to the quark-gluon plasma. Even close to the
 confinement temperature it is found to be in a good agreement with
 lattice data of several gauge systems.
 The model is applied to predict the equation of state of the quark-gluon
 plasma with realistic quark masses.
\end{abstract}

\pacs{PACS numbers: 11.10Wx, 12.38Cy, 12.38Mh}

\section{Introduction}\label{S:intro}

 Perturbation theory is a well established framework for approximative
 analytical calculations in quantum field theory.
 In numerous examples perturbative calculations have been shown to be
 successful in the description of processes and systems characterized
 by a weak coupling.
 For larger values of the coupling strength, however, extrapolated
 perturbative results do not yield appropriate approximations.
 So, e.\,g., the perturbative series of the deconfined QCD thermodynamical
 potential $\Omega$, which is known up to the order ${\cal O}(g^5)$
 \cite{ZK95}, shows even for moderate values of the coupling, say
 $g \gsim \frac12$, a strongly fluctuating behavior increasing with
 the order of the expansion, whereas lattice gauge simulations
 \cite{lat1,lat2} exhibit a rather smooth dependency $\Omega(g)$.

 This feature can be understood qualitatively by the asymptotic
 expansions $Z_n(g^2) = \sum_m^n z_m (g^2)^m$ of the function
 $
  Z(g^2)
  =
  \int_{-\infty}^\infty dx\, \exp\left\{-\frac12\, x^2 - g^2 x^4 \right\}
 $.
 This simple example \cite{IZ80} mimics the perturbative and full
 representation, respectively, of the partition function of an interacting
 field theory in zero dimension. (In fact, the coefficients $z_m$ count the
 number of diagrams contributing to the thermodynamical potential of the
 scalar $\phi^4$ theory considered below.)
 Although $Z(g^2)$ defined for complex $g^2$ possesses a cut along the
 negative axis (thus the series expansion is divergent), $Z_n$ still
 contains useful information as to be expected for perturbative results.
 The error of the approximation can easily be estimated to be
 $|Z(g^2)-Z_n(g^2)| \sim (n g^2)^n$, hence the optimal order to
 truncate the series expansion is $n^\star \sim 1/g^2$.

 These considerations generalized suggest that for the (at least qualitative)
 description of strongly coupled systems suitable lower-order perturbative
 approximations might be more appropriate than elaborated high-order
 calculations.
 For the QCD thermodynamical potential, this {\em leading-order conjecture}
 is also supported by the argument of minimal renormalization scale
 dependence of the truncated perturbative expansion.
 With this assumption for the regime of larger coupling strength, in Section
 \ref{S:phi4} the thermodynamical potential of the scalar $\phi^4$ theory is
 calculated in a selfconsistent nonperturbative approximation \cite{P98}.
 The emerging picture motivates a quasiparticle description of the
 thermodynamics of gauge systems as discussed in Section \ref{S:QCD}.
 The model is compared to available lattice data, and a prediction of
 the equation of state of the quark-gluon plasma (QGP) including strange
 flavor is given.

\section{Tadpole resummed $\phi^4$ theory}\label{S:phi4}

 It is well known that the (naive) perturbative calculation of the
 thermodynamical potential $\Omega$ of the massless scalar theory
 with the bare Lagrangian
 \be
  {\cal L}
  =
  \frac12\, (\pa_\mu \phi) (\pa^\mu \phi)
  - \frac{g_0^2}{4!}\, \phi^4
 \ee
 is plagued with infrared divergencies \cite{lB96}.
 These thermal divergencies can formally be treated by suitably rearranging
 the Lagrangian, e.\,g.\ by adding and subtracting a mass term to ${\cal L}$,
 yielding an equivalent theory with free massive propagators and an
 additional 2-point interaction.
 With such a reorganization of the perturbation theory, the expansion
 of $\Omega$ in the coupling strength, then well-defined up to 5th order,
 has been calculated \cite{PS95}.
 However, this series expansion exhibits a similar bad convergence as that
 of the QCD thermodynamical potential when being extrapolated to larger
 values of the coupling.
 As an alternative way to the perturbative expansion around the (modified)
 interaction-free limit, the Luttinger-Ward approach, originally derived
 for nonrelativistic fermion systems \cite{LW60} but readily generalized
 \cite{NC75}, is proposed here.
 Formulated in terms of full propagators, this formalism appears to be a
 more adequate description than perturbative calculations using free
 Green's functions.
 In the tadpole approximation considered below it leads to propagators with
 a mass-like selfenergy providing thus a less formal motivation of the
 perturbative reorganization technique.

 The Luttinger-Ward formulation of the thermodynamical potential of the
 scalar theory reads
 \bea
  \Omega
  &=&
  \frac12\, TV\sumint\,
  \left[ \ln (-\Delta^{-1}) + \Delta\Pi \right] \, + \, \Omega' \, ,
  \nonumber \\
  \Omega'
  &=&
  -\sum_m \frac1{4m}\, TV\sumint\, \Delta\Pi^{\rm s}_m
  \label{LW}
 \eea
 where $\Delta = (\Delta_0^{-1} - \Pi)^{-1}$ denotes the exact propagator
 and $\Pi$ is the exact selfenergy which can be decomposed into skeleton
 (i.\,e.\ 2-particle irreducible) contributions $\Pi_m^{\rm s}$ of order
 $m$. The sum-integral notation is specified below.
 The expression (\ref{LW}) obeys the fundamental condition \cite{LY60}
 \be
  \frac{\delta \Omega}{\delta \Pi}
  =
  0
  \label{stat}
 \ee
 which functionally relates microscopic properties ($\Pi$) to macroscopic
 quantities ($\Omega$) ensuring thermodynamical selfconsistency.
 Moreover, the representation (\ref{LW}) is an expedient starting point for
 a systematic (also referred to as symmetry conserving) approximation scheme.
 Approximating $\Omega'$ by the order-$n$ truncated series $\Omega_n'$ and
 the selfenergy by $\Pi_n = \sum_m^n \Pi_m^{\rm s}$ \rule[-1ex]{0ex}{1ex}
 yields an approximative expression $\Omega_n$ for the potential which
 strictly fulfills (\ref{stat}) \cite{B62}.

 Relying in the following on the conjecture stated in Section \ref{S:intro},
 $\Omega'$ is approximated in leading-loop order by
 $
  \Omega_1'
  =
  -3\unitlength1mm\begin{picture}(11,5)
      \multiput(4,1)(5,0){2}{\circle{5}}
      \multiput(4,1)(5,0){2}{\circle{4}}
    \end{picture}
 $ \rule[-2ex]{0ex}{2ex}
 where double lines represent the full propagator $\Delta_1 =
 (\Delta_0^{-1} -\Pi_1)^{-1}$ within the approximation considered.
 The corresponding selfenergy is given by
 \be
  \Pi_1
  =
  12\unitlength1mm\begin{picture}(10,5)
      \put(5,2){\circle{5}}
      \put(5,2){\circle{4}}
      \put(5,-0.5){\line(2,-1){4}}
      \put(5,-0.5){\line(-2,-1){4}}
    \end{picture}
  =
  12 \left(\frac{-g_0^2}{4!} \right)  T\sumint\, \Delta_1 \, .
  \label{Pi_1}
 \ee
 This implicit equation for $\Pi_1$ amounts to the summation of all super
 daisy selfenergy diagrams \cite{DJ74} of the (naive) perturbation theory.
 Due to the tadpole interaction process, $\Pi_1$ is real and momentum
 independent. Thus, $\Delta_1$ takes the form $\Delta_{m^2}(P) =
 (P^2-m^2)^{-1}$ of a free propagator with a mass term so the
 sum-integration in (\ref{Pi_1}) is elementary, though divergent.
 It can be regularized in $4-2\epsilon$ dimensions using in the imaginary
 time formalism
 $T\displaystyle\sumint
  =
  \left(\frac{e^\gamma\bar\mu^2}{4\pi} \right)^{\!\epsilon}
  T\sum_{p_0} \int\!\frac{d^{3-2\epsilon}p}{(2\pi)^{3-2\epsilon}}$,
 $p_0 = i\, 2\pi T\, n$, with $\gamma$ being Euler's number and $\bar\mu$
 the $\overline{\rm MS}$ renormalization scale,
 \bea
  T\sumint \Delta_{m^2}
  &=&
  \frac{m^2}{(4\pi)^2}
  \left[ \frac1\epsilon + \ln\frac{\bar\mu^2}{m^2} + 1 \right]
  \nonumber \\
  &&
  - \frac1{2\pi^2}\int_0^\infty dp\,
     \frac{p^2}{\omega_p}\, n_B(\omega_p/T)
  + {\cal O}(\epsilon) \, .
  \label{I}
 \eea
 Here the notation $\omega_p = (m^2+p^2)^{1/2}$ and the Bose function
 $n_B(x) = [\exp(x)-1]^{-1}$ are used.
 The first expression of the right hand side of equation (\ref{I})
 corresponds to the vacuum selfenergy term of a massive scalar particle.
 It contains a divergent contribution which in the present case cannot
 be compensated by a counter term since it is temperature dependent.
 However, both the divergent and the scale dependent term are absorbed when
 expressing $g_0$ by the physical coupling $g$ in equation (\ref{Pi_1}).
 To be consistent with the tadpole topology class for the selfenergy,
 $g$ is defined as the truncated vacuum scattering amplitude
 \unitlength1mm\begin{picture}(42,7)
   \put(5,1){\circle*{2}}
   \multiput(1,-3)(15,0){2}{\line(1,1){8}}
   \multiput(1,5)(15,0){2}{\line(1,-1){8}}
   \put(11.5,0){=}
   \put(26,0){+ 12}
   \put(39,1){\circle{5}}
   \put(39,1){\circle{4}}
   \put(39,3.5){\circle*{2}}
   \multiput(35,5.5)(4,-7){2}{\line(2,-1){4}}
   \multiput(35,-3.5)(4,7){2}{\line(2,1){4}}
 \end{picture} \rule[-3ex]{0ex}{3ex}
 at the momentum scale $s$:
 $
  g^2
  =
  g_0^2 - \frac12\, g^2 g_0^2 \, L(s)
 $
 with
 $
  L(s)
  =
  (4\pi)^{-2}
  \left[
    \epsilon^{-1} -\ln(s/\bar\mu^2) + 2 + {\cal O}(\epsilon)
  \right]
 $.
 With the identification $\Pi_1 = m^2$ and choosing $\sqrt s = eT$,
 the renormalized equation (\ref{Pi_1}) reads at $\epsilon \to 0$
 \bea
  m^2
  \;=\;
  \frac{g^2}2
  &&
  \left(
    \frac{m^2}{(4\pi)^2} \left[ \ln \frac{m^2}{T^2} - 1 \right]
  \right.
  \nonumber \\
  &&
  \left. \quad
   + \frac1{2\pi^2}\,
     \int_0^\infty dp\, \frac{p^2}{\omega_p}\, n_B(\omega_p/T)
  \right) .
  \label{m^2}
 \eea
 The solution $m^2(g)$ of this gap equation, when expanded in $g$,
 coincides with the perturbative selfenergy up to ${\cal O}(g^3)$.
 For larger values of $g$, in contrast to the perturbative result
 in the various known orders, it has a smooth behavior.

 In the tadpole approximation, the thermodynamic potential (\ref{LW})
 is given by
 \bea
  \Omega_1
  &=&
  \frac12\, TV\sumint\,
  \left[ \ln(-\Delta^{-1}_1) + \frac12\, m^2 \Delta_1 \right]
  \nonumber \\
  &=&
  \frac{TV}{2\pi^2}\, \int_0^\infty dp\, p^2
  \ln\left( 1 - \exp\{-\omega_p/T\} \right)
  \nonumber \\
  &&
  -\left[
     \frac{m^2 V}{8\pi^2}\,
       \int_0^\infty dp\, \frac{p^2}{\omega_p}\, n_B(\omega_p/T)
    +\frac{m^4 V}{128\pi^2}
   \right] ,
  \label{Omega1}
 \eea
 where (\ref{I}) and $\pa_{m^2} \ln(-\Delta^{-1}_1) = -\Delta_1$ were
 used for evaluating the sum-integral over $\ln(-\Delta^{-1}_1)$ up
 to an irrelevant integration constant.
 As to be expected for a consistent approximation, the divergent and
 renormalization scale dependent terms from the sum-integrals exactly
 cancel each other in $\Omega_1$.
 In analogy to the approximation $\Pi_1$ of the selfenergy, $\Omega_1$
 sums up all perturbative super daisy diagrams of the thermodynamical
 potential.

 It is emphasized that in equation (\ref{Omega1}) the first contribution
 is the thermodynamical potential of an ideal gas of quasiparticles with
 mass $m(T)$ as determined by the gap equation (\ref{m^2}).
 The remaining interaction contribution, when calculating the entropy
 $S_1 = -\pa\,\Omega_1 / \pa T$, compensates the temperature derivative
 terms of the quasiparticle mass. Thus, according to (\ref{stat}), $S_1$
 is given by the entropy of the ideal massive gas.

\begin{figure}
   \epsfxsize 7cm
   \centerline{\epsffile{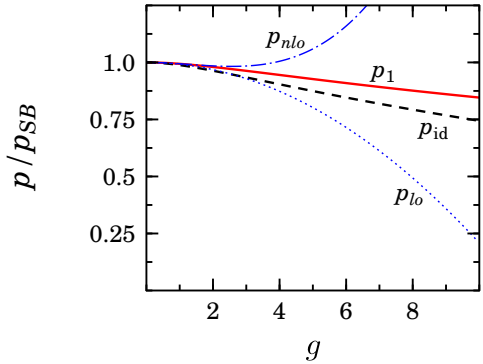}}
   \caption{The pressure $p_1$ and the ideal gas contribution $p_{\rm id}$
            as functions of the coupling strength $g$ in units of $p_{\rm SB}
            = \pi^2/90\, T^4$.
            For comparison, the leading and next-to-leading order
            perturbative results are shown as well.}
   \label{F1}
 \end{figure}

 Figure \ref{F1} shows the pressure $p_1 = -\Omega_1/V$ as a function
 of the coupling strength.
 In the range of the coupling considered, the positive interaction
 pressure turns out to be a correction to the ideal gas contribution
 $
   p_{\rm id}(T,m)
   =
   T/(2\pi^2) \int dp\, p^2 \ln( 1 - \exp\{-\omega_p/T\} )
 $
 at the 10\% level.
 For large values of $g$ the pressure $p_1$ does not deviate as much from
 the free limit as the naive extrapolation of perturbative results might
 indicate.
 This general feature is confirmed by other nonperturbative approaches.
 So the approximation $\Omega_1$ agrees with the result \cite{D98} derived
 by considering the O($N$) scalar theory in the limit $N \rightarrow \infty$.
 Without an expansion in $g$, the reorganized (screened) perturbation
 theory \cite{KPP97}, where a gap equation has to be taken as an `external
 information', leads to similar results.

 Compared to these approaches, the Luttinger-Ward {\em ansatz} is favored
 here because, by construction, it yields symmetry conserving approximations.
 Furthermore, expressing the thermodynamical potential in terms of
 full propagators, the formulation (\ref{LW}) presumably guarantees
 a well-organized asymptotic expansion \cite{RS97}.
 Finally, it is at least conceptionally straightforward to extend the
 selfconsistent resummation scheme to the next-to leading order where
 the functional $\Omega'$ is approximated by
 $
  \Omega_2'
  =
  \Omega_1'
  -12\unitlength1mm\begin{picture}(12,7)
     \put(6,1){\circle{8}}
     \put(6,1){\circle{7.7}}
     \put(6,1){\oval(7,3.5)}
     \put(6,1){\oval(6,2.7)}
  \end{picture} .
 $
 Due to the rising sun contribution
 \unitlength1mm\begin{picture}(18,7)
   \put(9,2){\circle{8}}
   \put(9,2){\circle{7.5}}
   \multiput(5,1.75)(0,0.5){2}{\line(1,0){8}}
   \put(5,2){\line(-1,-1){4}}
   \put(13,2){\line(1,-1){4}}
 \end{picture} \rule[-2ex]{0ex}{2ex}
 the selfenergy then becomes momentum dependent and complex thus the
 quasiparticles acquire a finite spectral width, a fact which requires
 more sophisticated techniques as those utilized here.

\section{A quasiparticle model for the QGP}\label{S:QCD}

 In QCD the difficulties mentioned at the end of the last section arise
 already at the 1-loop level.
 Therefore, in the following a more phenomenological approach, though
 closely related to the Luttinger-Ward formalism, is applied to describe
 the strongly coupled QGP.
 Temperatures closely above the confinement temperature $T_c \sim$ 200\,MeV,
 as they are possibly attainable in heavy-ion collisions, thereby require
 the appropriate treatment of the strange flavor. With a mass $m_{0s} \sim$
 150\,MeV the strange quarks represent a relevant non-light degree of
 freedom while the heavier flavors are suppressed.

 The plasma is characterized by broken Lorentz symmetry which leads to
 additional partonic excitations not existent in the vacuum.
 The effect of the symmetry breaking is yet marginal for excitations with
 momentum $p \gsim T$. Therefore, the hard `medium-bound' (longitudinal
 gluonic and helicity-flipped fermionic) modes are essentially not populated
 while the persisting excitations bear resemblance to the physical vacuum
 states.
 These thermodynamically relevant excitation are approximately described
 by a dispersion law $\omega^2(p) = m^2 + p^2$ with the effective gluon
 and quark masses given by \cite{lB96,P89}
 \bea
   m_g
   &=&
   \sqrt{ \frac16\left( N_c+\frac12 \, n_{\rm f} \right) g^2 T^2 } \, ,
   \nonumber \\
   m_q
   &=&
   \frac12
   \left(
     m_{0q}
   + \sqrt{ m_{0q}^2 + \frac{N_c^2-1}{2 N_c}\, g^2 T^2 }
   \right) .
  \label{m_eff}
 \eea
 These gauge invariant masses are generated dynamically and thus not in
 conflict with two propagating gluon excitations and with chiral symmetry
 for vanishing current quark masses $m_{0q}$.
 Within the mass parametrization (\ref{m_eff}) of interaction, sub-leading
 effects in $g$ like damping are neglected so the approach outlined below
 amounts to a partial resummation of relevant contributions similar to the
 Luttinger-Ward formalism.
 The expansion in $m/T \sim g$ yields the leading order perturbative QCD
 results \cite{P96}.

 The pressure of this system of effective quasiparticles with temperature
 dependent masses $m_i(T)$ is given reminiscently to (\ref{Omega1}) by
 \cite{GY95}
 \be
   p_{\rm qp}(T) = \sum_i p_{\rm id}(T, m_i(T)) - B(T) \, .
 \ee
 The function $B(T)$ resembles the interaction term in (\ref{Omega1}).
 It is not an independent quantity.
 From the thermodynamical relation (\ref{stat}), which in the present
 framework takes the form $\pa p_{\rm qp} / \pa m_i = 0$ (implying for
 chiral quarks $\langle \bar\psi_q \psi_q \rangle = 0$, thus the model
 respects chiral symmetry restoration in the plasma), $B$ is determined by
 \be
   \frac{dB}{dT}
   =
   \sum_i \frac{\pa p_{\rm id}}{\pa m_i}\, \frac{d m_i}{dT} \, .
  \label{dBdT}
 \ee
 As a special case, constant effective masses lead to $B(T) = const$
 and result in a bag model equation of state.

 The entropy density $s = \pa p / \pa T$ tests the occupation of the
 phase space, consequently $s_{\rm qp}(T) = \sum_i s_{\rm id}(T, m_i)$
 is insensitive to $B(T)$. This function has the obvious interpretation
 of the ground state energy of the quasiparticle system since the energy
 density $e = sT -p$ reads here
 $e_{\rm qp}(T) = \sum_i e_{\rm id}(T, m_i(T)) + B(T)$.

 To account for nonperturbative behavior near $T_c$, $g^2$ in (\ref{m_eff})
 is specified by the effective coupling
 \be
   G^2(T)
   =
   \frac{48 \pi^2}
        {(11\, N_c - 2\, n_{\rm f}) \,
         \ln[(T + T_s)/(T_c / \lambda)]^2}
  \label{G2}
 \ee
 where the parameter $T_s/T_c$ is a phenomenological regulator.
 For large temperatures, $G^2$ approaches the 1-loop running coupling
 with $T_c/\lambda \sim \Lambda_{\rm QCD}$.

 Equations (\ref{m_eff}) - (\ref{G2}) constitute the effective
 quasiparticle model which selfconsistently `continues' the leading
 order perturbative QCD thermodynamics.
 Corroborating the leading-order conjecture, it quantitatively reproduces
 thermodynamical lattice gauge data \cite{lat1,lat2} by suitably adjusting
 the parameters $\lambda$ and $T_s/T_c$ \cite{P96,K97}.
 Hereby it turns out that fixing the integration constant in
 (\ref{dBdT}) at the confinement temperature yields positive
 values $B_0 = \sum_i p_{\rm id}(T_c,m_i) - p_{\rm lattice}(T_c)$.
 This fact suggests a bag constant interpretation of $B_0$.
 In both the pure gauge system and the flavored plasma, however, at a
 temperature $T \sim 2 T_c$ the `bag' function $B(T)$ becomes negative
 due to the asymptotic behavior of the coupling which is opposed to
 the scalar model.

\begin{figure}
   \epsfxsize 7cm
   \centerline{\epsffile{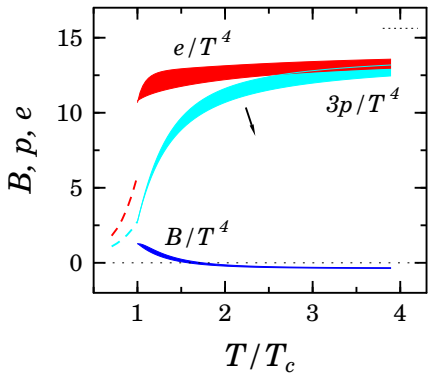}}
   \caption{The equation of state of the QGP with massive strange flavor as
            predicted by the quasiparticle model, and the function $B(T)$.
            Decreasing values of the model parameter $3 \le \lambda \le 10$
            are indicated by the arrow; the dotted short line marks the
            asymptotically free limit.
            The dashed lines represent pressure and energy density of the
            confined QCD phase approximated by a hadron resonance gas.}
   \label{F2}
 \end{figure}

 Motivated by the comparison to available lattice data, the quasiparticle
 model is now applied to predict the equation of state of the QGP with
 realistic quark masses which so far cannot be simulated numerically.
 The confinement temperature and the bag constant are chosen to be
 $T_c = $ 170\,MeV and $B_0 = $ (180\,MeV)$^4$.
 Equating the pressure of the confined phase (described by a hadron
 resonance gas) and $p_{\rm qp}$ at $T_c$ to fulfill Gibbs' condition
 relates the remaining two parameters $\lambda$ and $T_s/T_c$.
 In Figure \ref{F2} the predicted QGP pressure, energy density and the
 function $B(T)$ are shown for the large variation $3 \le \lambda \le 10$
 of the free parameter.
 Due to the stationary condition (\ref{stat}), the prediction is
 rather insensitive to the choice of $\lambda$, $T_c$ and $B_0$.
 Remarkably, for temperatures $T \gsim 2 T_c$ the energy density
 shows a saturation-like behavior at some 80-90\% of the free limit.
 It is worthwhile pointing out that the strange flavor yields an
 important contribution to thermodynamical quantities because the
 effective strange quark mass is yet smaller than the thermal gluon mass.

 In summary, a QGP quasiparticle model motivated by the Luttinger-Ward
 approach to $\phi^4$ thermodynamics is presented.
 Being successfully tested on lattice data of various gauge systems,
 it provides a realistic prediction of the equation of state of the QGP
 with strangeness. This allows further quantitative estimates of relevant
 observables in relativistic heavy-ion collisions.

\acknowledgments
 Interesting discussions with B.~K\"ampfer, G.~Soff and O.\,P.~Pavlenko
 are gratefully acknowledged.


\begin{references}

\bibitem{ZK95}
  C.~Zhai, B.~Kastening, Phys.~Rev.~D52 (1995) 7232

\bibitem{lat1}
  G.~Boyd, J.~Engels, F.~Karsch, E.~Laermann, C.~Legeland,
  M.~L\"utgemeier, B.~Petersson,
  Nucl.\ Phys.\ B469 (1996) 419,

\bibitem{lat2}
  J.~Engels, R.~Joswig, F.~Karsch, E.~Laermann, M.~L\"ut\-gemeier,
  B.~Petersson,
  Phys.\ Lett.\ B396 (1997) 210

\bibitem{IZ80}
 C.~Itzykson, J.\,B.~Zuber, {\em Quantum Field Theory},
 Mc\-Graw-Hill, New York (1980)

\bibitem{P98}
 A.~Peshier, B.~K\"ampfer, O.\,P.~Pavlenko, G.~Soff,
 Europhys.\ Lett.~43 (1998) 381

\bibitem{lB96}
 M.~Le Bellac, {\em Thermal Field Theory},
 Cambridge University Press, Cambridge (1996)

\bibitem{PS95}
  R.\,R.~Parwani, H.~Singh, Phys.\ Rev.\ D51 (1995) 4518

\bibitem{LW60}
  J.\,M.~Luttinger, J.\,C.~Ward, Phys.\ Rev.\ 118 (1960) 1417

\bibitem{NC75}
 R.\,E.~Norton, J.\,M.~Cornwall, Ann.\ Phys.\ 91 (1975) 106

\bibitem{LY60}
  T.\,D.~Lee, C.\,N.~Yang, Phys.\ Rev.\ 117 (1960) 22

\bibitem{B62}
  G.~Baym, Phys.\ Rev.\ 127 (1962) 1391

\bibitem{DJ74}
 L.~Dolan, R.~Jackiw, Phys.\ Rev.\ D9 (1974) 3320

\bibitem{D98}
  I.\,T.~Drummond, R.\,R.~Horgan, P.\,V.~Landshoff, A.~Reb\-han,
  Nucl.\ Phys.\ B524 (1998) 579

\bibitem{KPP97}
  F.~Karsch, A.~Patk\'os, P.~Petreczky, Phys.\ Lett.\ B401 (1997) 69

\bibitem{RS97}
  J.~Reinbach, H.~Schulz, Phys.\ Lett.\ B404 (1997) 291

\bibitem{P89}
 R.\,D.~Pisarski, Physica A158 (1989) 146

\bibitem{P96}
 A.~Peshier, B.~K\"ampfer, O.\,P.~Pavlenko, G.~Soff,
 Phys.\ Rev.\ D54 (1996) 2399

\bibitem{GY95}
 M.\,I.~Gorenstein, S.\,N.~Yang, Phys.\ Rev.\ D52 (1995) 5206

\bibitem{K97}
 B.~K\"ampfer, O.\,P.~Pavlenko, A.~Peshier, M.~Hentschel, G.~Soff,
 J.\ Phys.\ G23 (1997) 2001

\end{references}
\end{document}